\documentclass[conference]{IEEEtran}

\IEEEoverridecommandlockouts
\usepackage{bm}
\DeclareUnicodeCharacter{0301}{\'{e}} 
\usepackage{cite}
\usepackage{amsmath,amssymb,amsfonts}
\usepackage{algorithm2e} 
\usepackage{algpseudocode}
\usepackage{graphicx}
\usepackage{textcomp}
\usepackage{xcolor}
\usepackage{hyperref}
\usepackage{longtable}
\usepackage{array}
\usepackage{multirow}
\usepackage{soul}

\def\BibTeX{{\rm B\kern-.05em{\sc i\kern-.025em b}\kern-.08em
    T\kern-.1667em\lower.7ex\hbox{E}\kern-.125emX}}

\makeatletter
\def\ps@IEEEtitlepagestyle{%
  \def\@oddfoot{\mycopyrightnotice}%
  \def\@oddhead{\hbox{}\@IEEEheaderstyle\leftmark\hfil\thepage}\relax
  \def\@evenhead{\@IEEEheaderstyle\thepage\hfil\leftmark\hbox{}}\relax
  \def\@evenfoot{}%
}
\def\mycopyrightnotice{%
  \begin{minipage}{\textwidth}
  \centering \scriptsize
  Copyright~\copyright~2024 IEEE. Personal use of this material is permitted. Permission from IEEE must be obtained for all other uses, in any current or future media, including\\reprinting/republishing this material for advertising or promotional purposes, creating new collective works, for resale or redistribution to servers or lists, or reuse of any copyrighted component of this work in other works.
  \end{minipage}
}

\begin{document}

\title{Deep Learning with Uncertainty Quantification for
Predicting the Segmentation Dice Coefficient of
Prostate Cancer Biopsy Images\\
{\footnotesize \textsuperscript{}}
\thanks{}
}

\vspace{-6mm}
\author{\IEEEauthorblockN{1\textsuperscript{st} Audrey Xie}
\IEEEauthorblockA{\textit{Media Arts \& Sciences } \\
\textit{Massachusetts Institute of Tech.}\\
Cambridge, MA, USA \\
ahx@mit.edu}
\and

\IEEEauthorblockN{2\textsuperscript{nd}  Elhoucine Elfatimi}
\IEEEauthorblockA{\textit{ Pathology \& Lab Medicine} \\
\textit{University of California}\\
Irvine, CA, USA \\
eelfatim@uci.edu}
\and
\IEEEauthorblockN{3\textsuperscript{rd} Sambuddha Ghosal}
\IEEEauthorblockA{\textit{Media Arts \& Sciences } \\
\textit{Massachusetts Institute of Tech.}\\
Cambridge, MA, USA \\
sghosal@mit.edu}
\and
  \IEEEauthorblockN{4\textsuperscript{rth} Pratik Shah*\thanks{*Corresponding author: Dr. Pratik Shah Ph.D.(pratik.shah@uci.edu)}}
    \IEEEauthorblockA{\textit{ Pathology \& Lab Medicine} \\
    \textit{Biomedical Engineering} \\
    \textit{University of California}\\
    Irvine, CA, USA \\
    pratik.shah@uci.edu}
}

\maketitle

\vspace{-6mm}

%
\begin{abstract}
Deep learning models (DLMs) can achieve state-
of-the-art performance in histopathology image segmentation and classification, but have limited deployment potential in real-world clinical settings. Uncertainty estimates of DLMs can increase trust by identifying predictions and images that need
further review. Dice scores and coefficients (Dice) are benchmarks for evaluation of image segmentation performance, but usually not evaluated with DLM uncertainty quantification. This study reports DLM’s trained with uncertainty estimations, using randomly initialized weights and Monte Carlo dropout, to segment tumors from microscopic Hematoxylin and Eosin dye stained prostate core biopsy histology RGB images. Image level maps showed significant correlation [Spearman’s rank (p $<$ 0.05)] between overall and specific prostate tissue image sub-region
uncertainties with model performance estimations by Dice. This study reports that linear models that can predict Dice segmentation scores from multiple clinical sub-region based uncertainties of prostate cancer can be a more comprehensive performance
evaluation metric without loss in predictive capability of DLMs with a low root mean square error.\end{abstract}

\begin{IEEEkeywords}
 Deep learning, Prostate tumor segmentation, Dice scores, Uncertainty, Real-world deployment
\end{IEEEkeywords}

\section{Introduction}

\label{sec:intro}
Deep neural networks (DNNs) are increasingly used for image detection, classification, and segmentation in medical applications ~\cite{b1,b2}. However, DNNs are considered sub-optimal for clinical settings which require risk analysis and uncertainty estimation in model predictions~\cite{b3}. Many DNNs use softmax as the final activation function~\cite{b4}, but this   leads to incorrect interpretations of confidence levels~\cite{b5}. Bayesian neural networks offer an alternative by incorporating uncertainty into the model through a distribution over the weights~\cite{b6}, though they are computationally expensive and challenging to scale for high-dimensional images~\cite{b5}. A more feasible approach is using dropout to approximate Bayesian inference and minimize the Kullback-Leibler divergence between the approximate distribution and the posterior of a deep Gaussian process~\cite{b5}. Previous research has shown that optimizing small datasets with DNNs for medical image segmentation can lead to promising tool kits for clinical applications ~\cite{b52}. This work reports various uncertainty quantification methods to develop a comprehensive evaluation framework for prostate tumor segmentation models from  Hematoxylin and Eosin (H\&E)-stained histopathology images used in clinical diagnosis~\cite{b53,b54}.

This study leverages Monte Carlo dropout (MCD) and backpropagation for uncertainty quantification to improve the reliability of deep learning model (DLM) predictions in adenocarcinoma prostate biopsy images. It provides detailed uncertainty maps that correlate with segmentation performance metrics, such as Dice coefficients, offering a more clinically relevant evaluation of model accuracy. The approach highlights the importance of tumor region-based uncertainty assessments to potentially help clinicians make better-informed decisions, increasing likelihood of DLM integration into clinical workflows. Additionally, this study introduces a novel algorithm designed to calculate region-specific uncertainties, effectively predicting segmentation performance through linear models. The goal is to promote the adoption of uncertainty quantification methods, ensuring accurate and reliable histopathology image analysis across various clinical workflows.

\subsection{Related Work}

Uncertainty estimation techniques and metrics for deep neural networks in non-medical image classification using Monte-Carlo sampling have been reported ~\cite{b5}. However, uncertainty estimations, image processing methods, and Dice coefficient evaluations have not been fully adapted to digital histopathology, such as prostate biopsy image segmentation for tumor and non-tumor regions of interest (ROIs). Another approach for uncertainty quantification uses Normalizing Flows to model data probability distributions without MC dropout, relying on flow-based models instead ~\cite{b550}.

In medical applications, MCD has been used for classifying lung adenocarcinoma vs. squamous cell carcinoma without segmentation or Dice score evaluations of ROIs~\cite{b91}. Other studies trained MCD and Backpropagation-based DLMs using Dice-loss for tumor segmentation in oropharyngeal cancer PET/CT images~\cite{b92}, and developed a Dice-risk uncertainty measure that was outperformed by the coefficient of variation~\cite{b92}. However these studies did not calculate post hoc Dice scores for different ROIs, which is a focus of this paper. Additionally, a 3D U-Net using MCD generated uncertainty maps for lung cancer segmentation on CT images but did not evaluate Dice scores on prostate H\&E images~\cite{b93}. Linear regression models for predicting single Dice scores from grayscale CT scans based on image-level uncertainty have been reported ~\cite{b9a}. In a study on skin lesion classification, MCD-based uncertainty estimation led to segmentation errors, primarily due to class imbalances~\cite{b8, b13}. To our knowledge, no studies have reported the relationship between DLM uncertainties and the segmentation of different clinical ROIs in histopathology images of the prostate or other tissues using individual Dice coefficients. Prior works in tumor classification and segmentation using deep learning ~\cite{b10, b11, b12} could benefit from correlating uncertainty-based Dice scores with clinical outcomes, as done in this study.




\subsection{Summary of contributions}
In this study, a DNN was trained using MCD to calculate pixel-level uncertainty for accurate segmentation of prostate biopsy tumors from H\&E-stained RGB images. The model's performance was evaluated by five-fold cross-validation  and  performance metrics (see Methods and Table ~\ref{tab1}). Maximum (indicating the highest uncertainty), minimum (reflecting high confidence), mean (providing an overall measure of the model’s uncertainty across all pixels), and standard deviation (indicating the variability in uncertainty estimates) together provide a comprehensive view of prediction confidence (Table~\ref{tab1}).

The images were subdivided into three ROIs: tumor, non-tumor tissue, and non-tissue areas. Statistically significant correlations (p $<$ 0.05) between Dice scores and uncertainty were found by comparing the model's output with ground truth segmentation masks, and three individual region-based and overall uncertainty of each image were then calculated. A new algorithm (Algorithm  ~\ref{seg_algo1}) was proposed to calculate region-specific uncertainty, and linear models were developed to predict Dice scores using clinical region-specific uncertainties derived from the MCD model. These linear models use constants and uncertainty measures from the clinical ROIs, either individually or combined, to accurately predict model performance (Dice) (Eq.~\ref{equ2} and \ref{eq:1}).  This study reports that although image-level uncertainty can estimate Dice predictions (Eq.\ref{eq:3}), sub-clinical region-based uncertainties provide a more precise evaluation without losing predictive capabilities with low Root Mean Square Error (RMSE).
\begin{table}[t]
    \centering
    \caption{\footnotesize{Performance of a deep learning model augmented with Monte Carlo dropout for tumor segmentation from prostate core biopsy microscopic H\&E images, including AUROC - Area under the receiver operating characteristic, TPR – True positive rate, FPR – False positive rate, Std. dev. – Standard deviation.}}
    \resizebox{\columnwidth}{!}{
    \tiny
    \begin{tabular}{l l l l}
        \multicolumn{4}{c}{\textbf{Performance}} \\ \hline
        Dice & $0.8809$ & TPR & $0.7730$ \\
        AUROC & $0.8662$ & FPR & $0.0008$ \\
        \multicolumn{4}{c}{\textbf{Uncertainty}} \\
        Maximum & $0.0592$ & Mean & $0.0159$ \\
        Minimum & $0.0014$ & Std. dev. & $0.0130$ \\ \hline
    \end{tabular}
    }
    \label{tab1}
\end{table}

\section{Data description}

Prostate core biopsy images from Gleason2019 dataset (Grand Challenge for Pathology at MICCAI 2019)~\cite{b11} contains 244 tissue microarray RGB images (5120 $\times$ 5120 resolution) of H\&E-stained prostate tissue. Expert pathologists annotated each image with Gleason tumor grades (1-5) and generated Ground-truth (GT) segmentation masks. Gleason grades 1 and 2, which are rare, were labeled as non-tumor tissues (black pixels (class 0), while  Regions of Gleason Grades 3, 4, and 5 were considered tumors (white pixels (class 1)). The dataset includes 224 images with tumors and 20 without. A detailed description of data pre-processing and training of DLMs with five-fold cross-validation used in this study  can be found in~\cite{b10}.

\vspace{-1mm}
\section{Methods}
A Bayesian VGG-UNet CNN augmented with MCD was trained on prostate core biopsy images from the Grand Challenge for Pathology at MICCAI Gleason 2019 dataset~\cite{b11}. The network consists of an encoder-decoder structure~\cite{b10},  where the encoder  uses a modified version of the VGG-16 architecture without fully-connected layers~\cite{b12}. Uncertainty quantification was applied in the decoder to mitigate noise during the encoding process, with MCD used to approximate Bayesian inference by minimizing the posterior in a deep Gaussian Process~\cite{b5}. Additionally, Bayes-by-backprop (backprop) was employed to quantify uncertainty by evaluating the gradient of the loss function relative to the input images. Both MCD and backprop improved segmentation performance by identifying uncertain regions in the model’s predictions. Model training involved hyperparameter optimization, five-fold cross-validation, and advanced performance metrics to ensure a comprehensive evaluation of the segmentation results.



\subsection{Model training}
\begin{figure}[ht]
    \centering
    \includegraphics[width=0.48\textwidth, height=5cm]{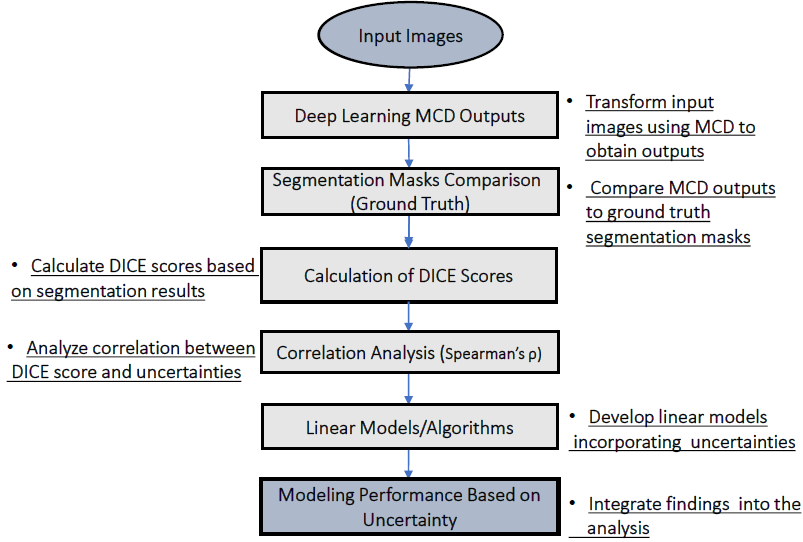} 
    \caption{\footnotesize{Integrated analysis flow: Sequential relationships in histopathology image analysis with deep learning, uncertainty maps, Dice score, and Monte Carlo
Dropout (MCD).}}
    \label{fig1}
\end{figure}

The prostate core biopsy data was split into 80\% for training ($\approx$  197 images) and 20\% for testing ($\approx$  47 images).  The model was trained for 50 epochs with a batch size of two, using the Adam optimizer with a learning rate set at $\alpha_{lr} = 0.001$. Input images were resized to 1024 $\times$ 1024 (W $\times$  H), while output images were set to 512 $\times$ 512. Model training and evaluation were performed on an NVIDIA GeForce RTX 3090 Ti with 24GB GPU memory. The workflow for model implementation and processing  with uncertainty integration approach is outlined in Fig.\ref{fig1}, starting with input images and applying the DLM MCD to obtain outputs.

The outputs were compared to ground truth masks, and Dice scores were calculated. Correlation analysis established the relationship between Dice scores and uncertainty estimations, leading to linear models for predictions. This workflow integrates segmentation performance of clinically relevant regions in histopathology images to assess overall model performance.

\subsection{Model prediction and evaluation}\label{sec_model_pred_eval}
In the process of evaluating the models predictive performance, occurrences of True Positives (TP), False Positives (FP), False Negatives (FN), and True Negatives (TN) across each image output were calculated. Specifically, a TP was recorded each time the model correctly identified a tumor pixel, whereas a TN was noted when it correctly identified pixels as non-tumor or non-tissue. Conversely, an FN occurred when tumor pixels present in the ground truth were overlooked by the model, and an FP was registered when pixels were incorrectly classified as tumor despite being non-tumor or non-tissue in the ground truth. The True Positive Rate (TPR) and False Positive Rate (FPR), alongside the Area Under the Receiver Operating Characteristic (AUROC) and the Dice Score (Dice) formed the basis for calculating key performance metrics. The Dice Score, a critical metric, assesses the congruence between the clinical ground truth labels and the model’s segmentation accuracy and was subsequently used to refine linear model estimations. The median values of Dice, AUROC, TPR, and FPR for the ensemble of 47 test images are documented in \textbf{(Table~\ref{tab1})}. Alongside these metrics, detailed analysis of pixel-level uncertainties for each image was perfomed. This involved conducting 50 Monte Carlo (MC) iterations to measure the uncertainty range, defined by the 67\textsuperscript{th} and 33\textsuperscript{rd} percentiles of the sigmoid output values of the model. These uncertainty values were then visualized through heatmaps (Fig.\ref{fig2}, and \ref{fig3}) and quantitatively captured in \textbf{(Table~\ref{tab1})}. From the MC data, the central tendencies for the mean uncertainties specific to each image were derived, leading to the calculation of an overall mean uncertainty for all test images. This comprehensive analysis revealed uncertainty levels ranging from a minimum of 0, indicating total certainty in predictions, to a maximum of 1, signifying complete uncertainty for further interpretation and  fine-tuning the model’s predictive reliability and accuracy. 

Mathematically, the model prediction for a given MCD model $M$ and input test image $X_{\text{test}_i}$ is calculated by:
\[ O_{M_p}^j = \text{argmax}(M_p(X_{\text{test}_i})) \]
for the $j$\textsuperscript{th} Monte Carlo (MC) iteration, where $M_p$ is the model $M$ in the testing phase. The final model output $\bm{O_{M_i}}$ is generated by averaging all 50 MC iteration outputs:
\begin{equation}
    \bm{O_{M_i}} = \frac{1}{\alpha} \sum_{j=1}^{\alpha} O_{M_p}^j
\end{equation}
where $\alpha = 50$ is the number of MC iterations. In the segmentation process, a value-based thresholding method is employed to determine the class to which each pixel belongs. A pixel is assigned to `class 0' (non-tumor), with a pixel value of `0' in the binary scale, if the mean value across all MC iterations is $\leq 0.95$. Conversely, it is assigned to `class 1' (tumor), with a pixel value of `1', if the mean value is $> 0.95$. Thus, the final pixel classification in $\bm{O_{M_i}}$ can be expressed as:
\[ \text{if } |p| > 0.95, \text{ then } |p| = 1 \text{ (tumor)}, \]
\[ \text{if } |p| \leq 0.95, \text{ then } |p| = 0 \text{ (non-tumor)}. \]
To construct the corresponding uncertainty map, $\bm{Unc_{map_i}}$, each MC iteration yields an uncertainty value defined by:
\[ Unc_{M_p}^j = \max(M_p(X_{\text{test}_i})). \]
The overall uncertainty map is then calculated as the difference between the 67\textsuperscript{th} and 33\textsuperscript{rd} percentiles of these values:
\begin{equation}\label{equ2}
    \bm{Unc_{map_i}} = \bm{P_{67}}(\{Unc_{M_p}^j\}_{j=1}^{\alpha}) - \bm{P_{33}}(\{Unc_{M_p}^j\}_{j=1}^{\alpha})
\end{equation}

\section{Results and discussion} \label{sec:results_and_disc}

\subsection{ Model Performance and Uncertainty Analysis} \label{sec:results_and_disc1}
\begin{figure}[h!]
\centering
\includegraphics[width=0.49\textwidth, height=0.22\textheight]{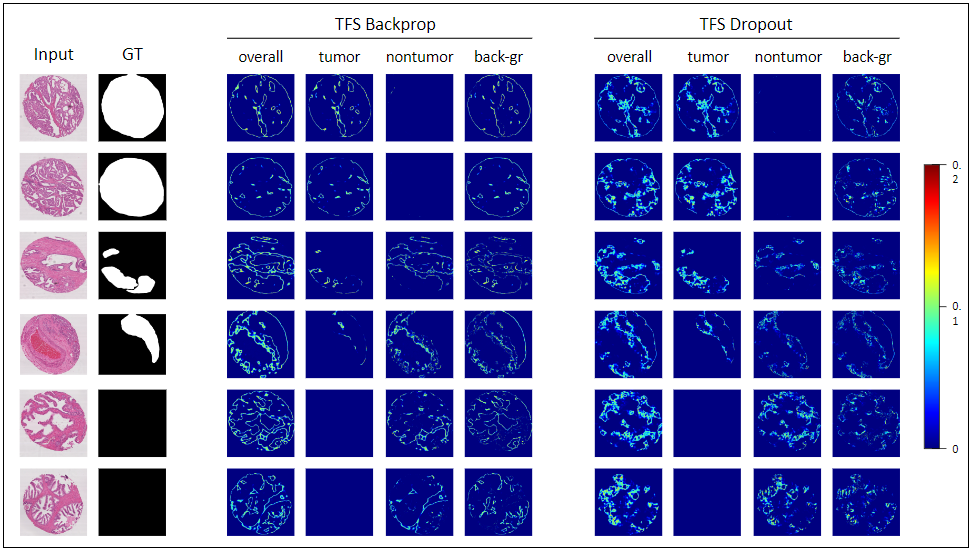}
\caption{\footnotesize{ Segmentation and Uncertainty Maps for Backprop and
Monte-Carlo Dropout Augmented Trained-from-Scratch (TFS) models.
“GT” – clinical ground-truth, “back-gr” – non-tissue regions. 
}}
\label{fig2}
\end{figure}

Five-fold cross-validation was used for segmentation of prostate tumors  and estimating segmentation uncertainties of the trained MCD model. Confidence intervals (CI) were calculated using the empirical bootstrap method with n = 5,000 simulations. Values for AUROC for all 47 test images were [0.8938 (95\% CI: 0.8582 - 0.9019)], Dice [0.8987 (95\% CI: 0.8561 -0.9095)], TPR [0.8382 (95\% CI: 0.7792 - 0.8980)] and FPR [0.0279 (95\% CI: 0.0296 -0.0829)]). The model in this study demonstrates superior generalization power, with a reported overall accuracy of 93.10\% and a loss of 0.2105. These results indicated that the MCD model achieved high performance (Table~\ref{tab1}) for prostate tumor segmentation. The overall uncertainty prediction were on the lower end (less than 1\%), with select individual ROI's within an input image showing uncertainties as high as 20\%. 
The segregated uncertainty maps for TFS models, augmented with MCD or backprop, are shown in Fig~\ref{fig2}.  The uncertainty maps are represented as heatmaps, deeper red indicates high uncertainty, and deeper blue indicates low uncertainty. These maps include binary segmentation masks and uncertainty estimates for tumor, non-tumor, and non-tissue regions. The overall uncertainty maps also show that tumor-containing images throughout the tissue regions  exhibited the least uncertainty, with  most of detected uncertainty at the tissue and non-tissue boundary regions Fig.~\ref{fig2}. Images without tumors showed higher false positive regions and greater non-tumor tissue uncertainty Fig.~\ref{fig2}. These visualizations can identify histopathology images with tumors by interpreting high segmentation uncertainties and the reliability of the model's output to inform clinical decision-making.
\vspace{-3mm}

A comparative analysis of backpropagation vs. dropout uncertainty was modeled with linear regression to predict the Dice Coefficient for histopathology image segmentation. The evaluation used metrics like the coefficient of determination ($R^2$), significance tests, and RMSE. While backpropagation showed a higher $R^2$ value (0.502049), indicating a superior overall fit, the uncertainty dropout model had better predictive accuracy with a lower RMSE (0.062141 vs. 0.07486). RMSE was prioritized in this study for its emphasis on accurate predictions and nuanced uncertainty modeling, The RMSE metric identified the linear regression model between uncertainty dropout  and Dice scores for its superior predictive accuracy and advanced uncertainty modeling, aligning with the study's focus on precision in predictions.


\vspace{-1mm}
\subsection{Qualitative evaluation} \label{sec:results_and_disc2}
\begin{figure}[h!]
\centering
\includegraphics[width=0.48\textwidth, height=0.22\textheight]{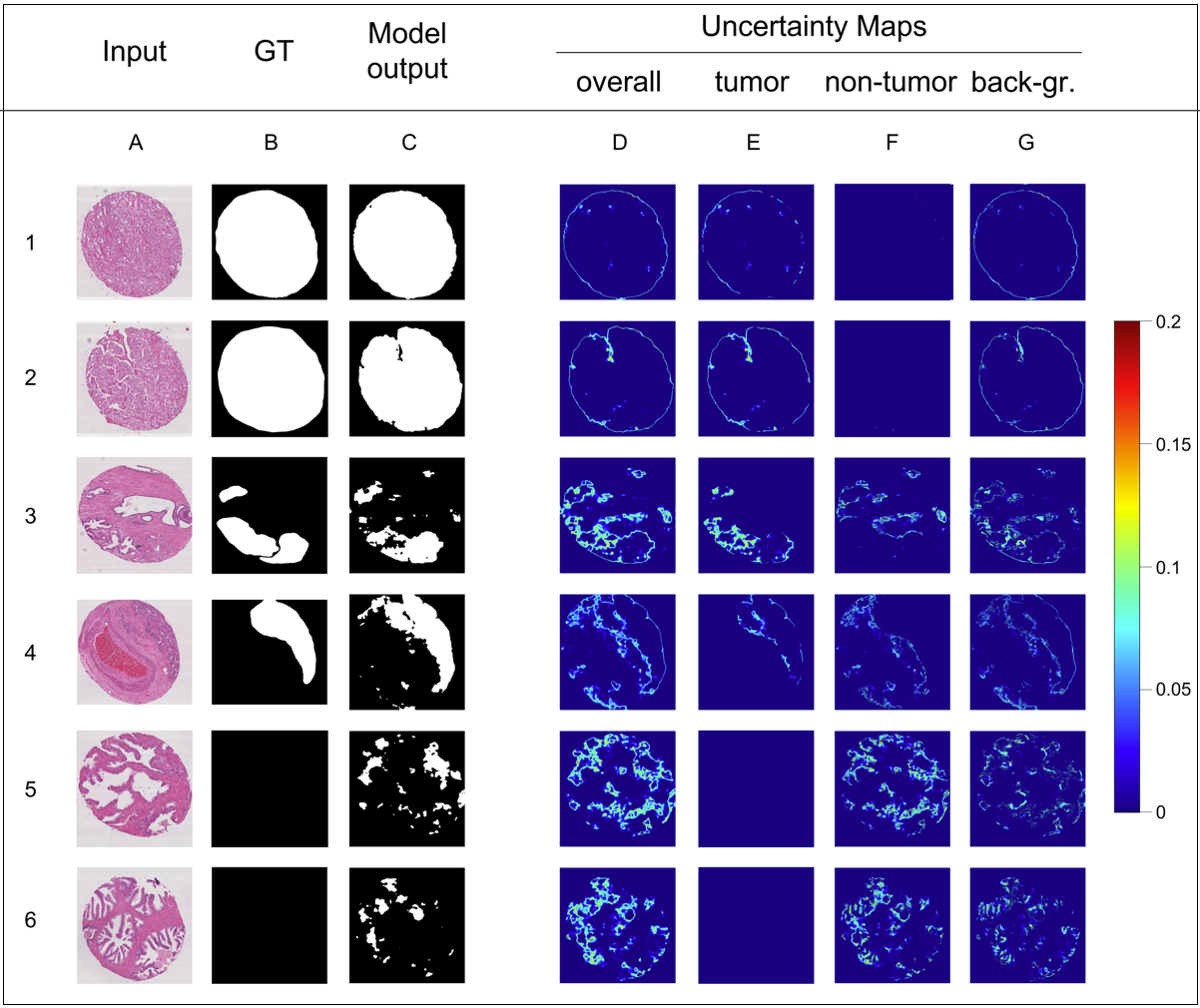}
\caption{\footnotesize{Visualization of model output segmentation and region-based uncertainty maps for prostate tumor segmentation using a Monte Carlo dropout (MCD) deep learning model trained on Hematoxylin and Eosin (H\&E) stained prostate biopsy images. \textbf{Columns left to right:} \textbf{A}- Input RGB image; \textbf{B}- Clinical ground-truth binary mask; \textbf{C}- MCD output segmentation binary mask; \textbf{D}- MCD model uncertainty estimate maps; \textbf{E}- Tumor tissue uncertainty map; \textbf{F}- Non-tumor tissue uncertainty map; \textbf{G}- Non-tissue uncertainty map. Color bar shows model uncertainty, with red for higher and blue for lower uncertainty (minimum value 0). ``GT" -- clinical ground-truth, ``back-gr" -- non-tissue regions.}}
\label{fig3}
\end{figure}

\textbf{Figure~\ref{fig3}} shows four tumor-containing and two non-tumor prostate images from the test data and MCD model output segmentation masks and corresponding uncertainty heatmaps from the trained model. Model predictions $\bm{O_{M_i}}$ and overall uncertainty map, $\bm{Unc_{map_i}}$ obtained using step 2 of Algorithm 1 are shown in columns C and D, respectively. The region-based uncertainty maps generated using step 2 of Algorithm 1, $\bm{Unc_{i_{T}}}$, $\bm{Unc_{i_{NT}}}$, and $\bm{Unc_{i_{NTi}}}$ are shown in columns E, F, and G, respectively. Images 1 to 4 contained tumors and images 5 and 6 did not contain ground-truth tumor signatures. Predictions for images 1 and 2 exhibited negligible false positive or false negative regions, images 3 and 4 showed low false positives and good segmentation accuracy (Fig.~\ref{fig3}). Images 5 and 6 exhibited the highest false positive regions as related instances were less frequent in the training data compared to images that had tumors in them (Fig.~\ref{fig3}). Overall uncertainty maps (column D) showed that images with tumors all throughout the tissue regions (images 1 and 2) exhibited the least uncertainty (Fig.~\ref{fig3}). Majority of the detected uncertainty for these images were at the tissue and non-tissue boundary regions. The MCD model showed no uncertainties in the non-tissue regions except for the junction between tissue and non-tissue regions (Fig.~\ref{fig3}). This indicated that the model has learned the non-tissue region well and is uncertain either at the non-tissue and tissue junctions, and more frequently at the non-tumor tissue and tumor tissue region intersections.

This was especially true for  regions indicated by the ground-truth as containing tumorous tissues in images 2 and 3 in {figure~\ref{fig3}}. For images 5 and 6 the absence of tumors gives the tissue island like appearance which may lead to a greater degree of confusion for the model and non-tumor tissue uncertainty. Visualizing these individual ROI-based uncertainty maps (Fig.~\ref{fig3}), also helped in identifying regions of high uncertainties and make informed decisions on whether to discard a model's output or not based on where the higher uncertainties arose from. For example in image 3 in Fig.~\ref{fig3}, high uncertainties were observed in the tumor region indicating that the model may show poor performance within tumor regions in similar images during inference, while similar uncertainties were observed in image 2 for the tumor and non-tissue regions indicating that predictions may be acceptable for cases when differential diagnosis depended on disease pixels. 

\vspace{-2mm}
\subsection{Performance Modeling Through Uncertainty Quantification} \label{sec:results_and_disc3}

\begin{algorithm}[t!]
\small
\caption{\textbf{ROI-Unc}: Modelling deep learning performance with clinical-class based image region specific uncertainty}\label{seg_algo1}
\hrule
\vspace{1mm}
\textit{\textbf{Step 1: -Model predictions and associated uncertainty maps}}
\\
\textbf{Input:} Trained MCD model $M$; Test data $X_{test}$ with $n$ images and ground truth $X_{GT}$. 
\newline
\textbf{Output:} Segmentation output, $\bm{O_{M_i}}$ and uncertainty map, $\bm{Unc_{map_i}}$ for each $X_{test_i}$.
\newline
Thus, $\forall$ i = [1, 2,....., $n$] images, 
\[
X_{Dice_i} = Dice(O_{M_i}, X_{GT_i}), List_{Dice} = [X_{Dice_i}]_{i=1}^{n}=
\]
\[
List_{Unc} = [Unc_{map_i}]_{i=1}^{n} 
\]

\textit{\textbf{Step 2: Generate ROI uncertainty:}}
\\
\textbf{Input:} Given a test image, $X_{test_i}$ and a ground-truth mask, $X_{GT_i}$, the following were generated: 
\begin{itemize}
    \item $X_{test_i}$-tumor = $X_{GT_i}$, where 0 represents non-tissue and 1 (255 in grayscale) represents tissue regions with or without tumors.
    \item $X_{test_i}$-non-tumor = $X_{test_i}$-tumor$^{C}$, complement of $X_{test_i}$-bin, a binary image where 1 is non-tissue and 0 is tissue.
\end{itemize}

\textbf{Output:} Uncertainty maps corresponding to these regions are generated as follows:
\begin{itemize}
    \item $\bm{Unc_{i_L}}$ = $Unc_{test_i}$-tumor = ($X_{test_i}$-tumor $\odot$ $Unc_{map_i}$); $\bm{List_{T_{unc}}} = [mean(Unc_{i_L})]_{i=1}^{n}$.
    \item $\bm{Unc_{i_NL}}$ = $Unc_{test_i}$-non-tumor = ($X_{test_i}$-non-tumor $\odot$ $Unc_{map_i}$); $\bm{List_{NT_{unc}}} = [mean(Unc_{i_NL})]_{i=1}^{n}$. The symbol $\odot$ denotes the Hadamard product.
\end{itemize}

\textit{\textbf{Step 3: Generate linear regression models:}} 
\\
\textbf{Input:} Dependent variable, $List_{Dice}$ = $Y_{Dice}$ (say) $= [y_{Dice_i}]_{i=1}^{n}$ for $n$ images, and independent variables:
\begin{itemize}
    \item $List_{Unc}$ = $X_0$ (say) $= [x_{0i}]_{i=1}^{n}$ (overall uncertainty maps as presented in Fig. ~\ref{fig3} column D),
    \item $List_{T_{unc}}$ = $X_1$ (say) $= [x_{1i}]_{i=1}^{n}$ (tumor tissue uncertainty maps as presented in Fig. ~\ref{fig3} column E),
    \item $List_{NT_{unc}}$ = $X_2$ (say) $= [x_{2i}]_{i=1}^{n}$ (non-tumor tissue uncertainty maps as presented in Fig. ~\ref{fig3} column F) and
    \item $List_{NT_{unc}}$ = $X_3$ (say) $= [x_{2i}]_{i=1}^{n}$ (uncertainty maps for non-tissue pixels without tissue as shown in Fig. ~\ref{fig3} column G).
\end{itemize}
These predict $Y_{Dice}$, where $n$ is the total number of images under consideration.

\textbf{Output:} Model $Y_{Dice}$ in terms of $X_1$, $X_2$, and $X_3$ as follows:
\begin{equation} \label{eq:1}
    Y_{Dice} = \alpha_0 + \alpha_1*X_1 + \alpha_2*X_2 + \alpha_3*X_3
\end{equation}

For individual ROIs:
\begin{subequations}\label{eq:44}
\begin{align}
  Y_{Dice} &= \beta_0 + \beta_1 X_1 \tag{4(i)}\label{eq:4i} \\
  Y_{Dice} &= \gamma_0 + \gamma_1 X_2 \tag{4(ii)}\label{eq:4ii} \\
  Y_{Dice} &= \delta_0 + \delta_1 X_3  \tag{4(iii)}\label{eq:4iii}
\end{align}
\end{subequations}

And for the overall uncertainty, $X_0$:
\begin{equation} \label{eq:3}
    Y_{Dice} = \theta_a + \theta_b X_0
\end{equation}

\hrule
\end{algorithm}

This study proposes Algorithm~\ref{seg_algo1} to calculate estimated performance (Dice) of a model using uncertainties for clinical ROIs in a medical image. Eq.~\ref{eq:1} calculated the Dice of the segmentation model based on a linear combination of tumor-tissue, nontumor tissue and non-tissue uncertainties, and equations \ref{eq:4i}, \ref{eq:4ii},  and \ref{eq:4iii}, calculated the Dice of the segmentation model  from the three individual ROIs. Two subsets of test data were used for calculating linear models and for obtaining the values of the coefficients for equations  ~\ref{eq:1} to~\ref{eq:3} and are described below:

\subsubsection{Tumor-containing images} \label{sec:results_and_disc3a}
Forty four images that contained tumor labels from the test data set were used as the first subset. From these tumor containing
images, we obtained  $\alpha_0$ = 1.0036, $\alpha_1$ = -12.2397,  $\alpha_2$ = -19.5844 , and $\alpha_3$ = -6.2364; for tumor uncertainty (Spearman's correlation coefficient ($\rho$) = -0.4878), $\beta_0$ = 0.9264 and $\beta_1$ = -7.9295; for non-tumor uncertainty ($\rho$ = -0.5012),    $\gamma_0$ = 0.9011 and $\gamma_1$ = -13.9173, for non-tissue uncertainty
 ($\rho$ = -0.5969), $\delta_0$ = 0.9702 and $\delta_1$ = -26.9116,  and for the overall
uncertainty ($\rho$ = -0.8496), $\theta_a$ = 1.0074 and $\theta_b$ = -14.9116. Low RMSE of $\leq 0.1$ was calculated for all four linear models for the tumor containing images.
 
 \subsubsection{ Tumor-free images} \label{sec:results_and_disc3b}
Three images from the test data and 17 images from the training
data that did not contain any tumor labels were merged
into the second subset. From these tumor-free images  $\alpha_0$ = 0.9991, $\alpha_1$ = 0 (which is expected since the tumor regions did not contribute to the model uncertainties or performance),  $\alpha_2$ = -12.306 and $\alpha_3$ =7.2247; for  non-tumor uncertainty (Spearman's correlation coefficient ($\rho$ = -0.8496), $\gamma_0$ = 0.9993 and $\gamma_1$ = -8.5907; for non-tissue uncertainty ($\rho$ = -0.8496), $\delta_0$ = 0.9993 and $\delta_1$ = -16.02, and for the overall uncertainty ($\rho$ = -0.8496), $\theta_a$ = 0.9994 and $\theta_b$ = -5.6296 were calculated. The RMSE values for all four linear models were $\leq 0.16$. 
The RMSE values for the tumor-free images
for four linear models were even lower at 0.014. For all
derived coefficients, except for $\alpha_3$ for the tumor-free images,
negative coefficient values in the majority of the calculations indicated that higher uncertainty in the MCD model predictions decreased its performance.
From the coefficients derived for Eq.~\ref{eq:1}, the significance of each clinical region in determining overall model performance becomes clear. A more negative coefficient implies a greater reduction in Dice scores. For tumor-containing images, non-tumor tissue region uncertainties had the highest coefficient value, meaning they had the largest negative impact on performance. The MCD model was most affected by non-tumor tissue uncertainty, followed by tumor-tissue regions, with non-tissue pixels contributing about half as much. In tumor-free images, non-tumor tissue uncertainty played a more substantial role than pixels without tissue, as indicated by the positive $\alpha_3$ coefficient.

\vspace{-2mm}

\section{Discussion}

Uncertainties in DLMs can significantly impact predictions, especially in critical ROIs such as tumor, non-tumor, and non-tissue areas. These uncertainties are crucial for model reliability in clinical settings. In this study, MCD and backpropagation were used to quantify uncertainties, with Dice scores evaluating segmentation accuracy (Figs.\ref{fig2} and \ref{fig3}). Tumor regions exhibited lower uncertainties, while higher uncertainties occurred at tissue boundaries. Non-tumor images showed more false-positive uncertainties in non-tumor regions, demonstrating the method's potential for real-world applications, even in scenarios where ground truth may be limited, as outlined in Eq.\ref{eq:3}). A new approach (Algorithm ~\ref{seg_algo1}) was developed to calculate region-specific uncertainties and predict performance. The non-tissue region ($X_3$) contributed to overall uncertainty but had less effect on the Dice score than non-tumor tissue regions. The prediction of model performance, based on Dice scores, was guided by overall uncertainty, as described in Equation ~\ref{eq:3}. Backpropagation, combined with MCD, offered deeper insights into areas of low prediction confidence, supporting more informed clinical decision-making. Future research could apply these uncertainty quantification techniques to other medical imaging datasets for greater model reliability.

\vspace{-1mm}
\section{Conclusion and Limitations}

This study demonstrates that region-based linear models can predict deep learning performance (Dice) using clinically meaningful uncertainty estimates (Equations \ref{eq:1} -\ref{eq:3}), Previous studies typically focused on either Dice or overall uncertainty, but this work uniquely combines both, offering a more comprehensive evaluation of model performance, especially for prostate adenocarcinoma segmentation. To our knowledge, this is the first time, three distinct clinical region-based uncertainties—tumor, non-tumor, and non-tissue regions—were decoupled and analyzed; showing that linear models accurately predict performance across these regions. Spearman’s rank correlation ($p < 0.05$) revealed a strong negative correlation between increased uncertainty and lower Dice scores. Algorithm ~\ref{seg_algo1} was introduced for unsupervised performance estimation, allowing prediction of Dice scores and generation of uncertainty maps, aiding in clinical decision-making.Although the study is currently limited to the prostate biopsy dataset, MCD, and Bayes-by-backprop techniques, it lays the groundwork for applying these uncertainty estimation methods to other datasets and clinical contexts. Future research should expand on these findings, exploring broader applications to enhance the reliability and accuracy of deep learning models in healthcare.

\bibliography{refs}  
\section{Data, Code and Model Availability}

All resources, results, code, models, and data from this study are available on GitHub \url{https://github.com/Prof-Pratik-Shah-Lab-UCI/Prostate_Project_2024} upon request to the corresponding author.

\end{document}